# Spatially and spectrally resolved orbital angular momentum interactions in plasmonic vortex generators

*Jordan A. Hachtel[1,2,3*], Sang Yeon Cho[4*], Roderick B. Davidson II[2], Matthew F. Chisholm[3], Richard F. Haglund[2], Juan Carlos Idrobo[1], Sokrates T. Pantelides[2,3,5], Benjamin J. Lawrie[6]*

[1] Center for Nanophase Materials Science, Oak Ridge National Laboratory, Oak Ridge, TN 37831 USA
[2] Department of Physics and Astronomy, Vanderbilt University, Nashville, TN 37235-1807 USA
[3] Materials Science and Technology Division, Oak Ridge National Laboratory, Oak Ridge, TN 37831 USA
[4] Klipsch School of Electrical and Computer Engineering, New Mexico State University, NM, 88003 USA
[5] Department of Electrical Engineering and Computer Science, Vanderbilt University Nashville, TN 37235 USA
[6] Computational Sciences and Engineering Division, Oak Ridge National Laboratory, Oak Ridge, TN 37831 USA

*This manuscript has been authored by UT-Battelle, LLC under Contract No. DE-AC05-00OR22725 with the U.S. Department of Energy. The United States Government retains and the publisher, by accepting the article for publication, acknowledges that the United States Government retains a non-exclusive, paid-up, irrevocable, worldwide license to publish or reproduce the published form of this manuscript, or allow others to do so, for United States Government purposes. The Department of Energy will provide public access to these results of federally sponsored research in accordance with the DOE Public Access Plan (*http://energy.gov/downloads/doe-public-access-plan*).*

## Abstract

Understanding the near-field electromagnetic interactions that produce optical orbital angular momentum (OAM) is central to the integration of twisted light into nanotechnology. Here, we examine the cathodoluminescence (CL) of plasmonic vortices carrying OAM generated in spiral nanostructures through scanning transmission electron microscopy (STEM). The nanospiral geometry defines the photonic local density of states (LDOS) sampled by STEM-CL, which provides access to the phase and amplitude of the plasmonic vortex with nanometer spatial and meV spectral resolution. We map the full spectral dispersion of the plasmonic vortex in the spiral structure and examine the effects of increasing topological charge on the plasmon phase and amplitude in the detected CL signal. The vortex is mapped in CL over a broad spectral range, and deviations between the predicted and detected positions of near-field optical signatures of as much as 5 *per cent* are observed. Finally, enhanced luminescence is observed from concentric spirals of like handedness compared to that from concentric spirals of opposite handedness, indicating the potential to couple plasmonic vortices to chiral nanostructures for sensitive detection and manipulation of optical OAM.

## *Main Text*

**Introduction**

Light possessing orbital angular momentum (OAM) is a topic of widespread current interest[1–4]. The helical wavefront of optical OAM modes can impart angular momentum to micro- and nanoscale structures and provide novel routes to optical manipulation and trapping[5,6]. Additionally, the added degrees of freedom provided by an orthonormal angular momentum basis set enables large-scale multiplexing of both classical and quantum communications, and provides a framework for emerging studies in quantum information science[7–11]. Furthermore, it has been proposed that OAM modes can potentially be used to detect molecular chirality[12–15], which plays a dominant role in biological and chemical processes[16]. Many of the above applications are centered in nanotechnology, which has made the integration of optical OAM to nanoscale devices a focal point of current research interests[17–19].

Surface plasmons are routinely used to manipulate light at the nanoscale, driving the development of many kinds of metasurfaces and asymmetric plasmonic nanostructures that enable the on-chip generation and control of OAM[20–25]. Nanoscale OAM effects in plasmonic nanostructures have been observed by scanning near-field techniques[26–29] and time-resolved photoemission electron microscopy (PEEM)[30], highlighting the broad potential for near-field applications of OAM modes in nanotechnology. However, these previous characterizations of OAM in plasmonics have all been performed with monochromatic, resonant optical excitation based on predicted discrete wavelengths for the topological charges defined by the plasmonic structures. Thus, a nanoscale description of dispersion in plasmonic vortices with high spectral resolution is a critical step in developing proposed advanced applications for nanophotonic OAM states[31].

In this paper, we employ cathodoluminescence (CL) spectroscopy to examine plasmonic vortices generated in spiral nanostructures using scanning transmission electron microscopy (STEM). Archimedean spiral channels present an ideal testbed for the near-field study of OAM because they can be designed to produce plasmonic vortices with geometry-dependent OAM in structures of sub-micron dimensions. The fast electrons of the STEM initiate a sub-femtosecond impulse excitation localized to a spot size of ~1 nm diameter, providing access to the photonic local density of states (LDOS) of the system. Furthermore, STEM-CL can be used to simultaneously examine the near-field plasmonic phase and amplitude across a broad spectral range with high spatial resolution, revealing the effects of localized inhomogeneities on the optical properties of the structure. As a result, coupling between plasmonic vortices and additional chiral nanostructures can be observed through enhanced CL in systems where both the vortex generator and the secondary nanostructure have the same handedness, enabling direct detection of spatially- and spectrally-resolved OAM interactions.

**<u>Materials and Methods</u>**

*Archimedean Spiral Plasmonic Vortex Generators*

Simulations have previously shown that plasmonic vortices possess optical OAM modes in Archimedean nanospiral grooves via interference between the exciting circularly polarized light beam and the scattered surface plasmon polaritons (SPPs) in the system. The measured OAM combines coherently the angular momentum of the exciting source with the plasmonic response of the resulting SPP vortex[20]. Here, by studying the spiral with an exciting electron beam carrying zero angular momentum, the OAM of the plasmonic vortex can be accessed directly.

The Archimedean spiral is described by the polar equation $r(\theta)=r_0+d\cdot\theta$, where $r_0$ is the initial radius, $\theta$ is an azimuthal coordinate, and $d$ is the interarm distance. The spirals are fabricated by evaporating 100 nm of Ag onto an electron-transparent SiN membrane (~50 nm thick), followed by focused ion beam (FIB) milling of the spiral channel in the Ag. The channel width is chosen to be 100 nm to avoid Ag re-deposition during FIB milling; more details about the channel width are included in the Supplementary Information.

Each point along the edge of the channel serves as a scattering source, creating surface plasmon polaritons (SPPs) that propagate in all in-plane directions. If the dimensions of the spiral are chosen such that the arm spacing, $d$, is an integer multiple of the SPP wavelength, $\lambda_{SPP}$, then there is a $2m\pi$ phase offset between SPPs generated at beginning and end of the channel, where $m$ is the integer multiple. The interference between all SPPs generated inside the nanospiral channel arms results in a composite plasmonic response describable as a Bessel mode[20]. The electromagnetic field of the mode in cylindrical coordinates is described by

$$U_\ell(r,\theta,z) = e^{i\ell\theta}J_\ell(k_r r)e^{ik_z z}, \qquad (1)$$

where $k_z$ and $k_r$ are the vertical and radial wavenumbers, and $J_\ell$ is an $\ell^{th}$-order Bessel function of the first kind. The order, or topological charge, of the plasmonic vortex in the spiral channel system is determined by the vector sum of the angular momentum in the exciting source, and the topological charge produced by the nanostructure geometry. The electron beam carries zero angular momentum and is properly described by a $\ell=0$ Bessel mode; therefore the OAM of the system is determined entirely by the geometry of the spiral. The topological charge of the structure is thus defined by the ratio between the inter-arm separation and the SPP wavelength, and can be written as $\ell=d/\lambda_{SPP}$.

*Cathodoluminescence in the Scanning Transmission Electron Microscope*

To access the full broadband optical response of the spiral structure with nanometer-scale-spatial resolution, a hyperspectral map of the spiral channel is obtained using STEM-CL by means of a spectrum image (SI). The SI is a three-dimensional data set that describes the visible to near-infrared spectrum of the CL at each pixel sampled by the rastered electron beam. From the SI, slices of the hyperspectral response are generated in which the integrated intensity over a specific spectral range is plotted *at each pixel* to generate 2D maps of specific resonant responses. The SI shown in this paper is acquired at a dwell time of 2 seconds per pixel, which limits the number of pixels that can be acquired for the SI, so band-pass (BP) filtered CL is used to enable acquisition at higher spatial resolution (55x61 pixels for the SI, 256x256 for BP-CL). In BP-filtered CL, the output luminescence is collected by a photomultiplier tube (PMT) detector which yields improved signal-to-noise ratios and faster imaging at the expense of inherent spectral resolution.

## Results and Discussion

### *Spatially and Spectrally Resolved Plasmonic Vortices in Cathodoluminescence*

The first spiral geometry examined is illustrated schematically in Figure 1a. The spiral dimensions are chosen such that $r_0$ and $d$ are both equal to $\lambda_{SPP}$, which is calculated to be 630 nm for the vacuum/Ag/SiN multilayer structure. The nominal designed dimensions of the spiral channel yield a free space wavelength for the OAM mode at 660 nm, and since the inter-arm spacing is equal to the SPP wavelength the topological charge for the plasmonic vortex is equal to 1. Figure 1b shows a bright-field (BF) STEM image of a spiral channel, with a red rectangle defining the region of interest for the SI. Figure 1c shows a frame-by-frame picture of the SI in 10 nm wavelength bins at 20 nm pitch from 440 nm to 740 nm, illustrating the combined spatio-spectral response of the plasmonic vortex generator.

At both short- and long-wavelength extremes of the SI, the CL illustrated in Figure 1c does not exhibit a vortex response and is dominated by linear interference fringes. However, over the middle spectral range (approximately 540 nm-680 nm), spiral arms with high CL intensity emerge and the plasmonic vortex is resolved. A video containing the plasmonic response of the system over the full collected spectral range is included in the Supplementary Information and shows the vortex dispersion with a large amount of narrow wavelength bins (~4 nm per bin).

The near-field response of the plasmonic vortex under electron-beam excitation can be modeled by finite-difference time-domain (FDTD) simulations, in which the fast electrons are treated as point dipoles oriented along the beam direction ($z$-direction)[32]. Figure 2a and 2b show the phase (a) and amplitude (b) of the plasmonic vortex from a $z$-oriented dipole excitation at the origin of the spiral at a wavelength of 660 nm. The near-field signature of a $\ell=1$ Bessel mode is observed in the dipole simulations, comprising a vortex singularity at the origin in the plasmon phase plot and rotationally-symmetric concentric rings of high intensity centered at the origin. However, comparing the near-field simulations to the detected CL response at 660 nm (shown in Figure 2c) reveals that neither the near-field amplitude nor phase qualitatively reproduces the CL. This critical point is a result of the fact that, unlike previous near-field characterization of plasmonic vortices, CL relies on local excitation with global detection.

In STEM-CL, a converged electron probe (diameter less than 1 nm) is rastered across the sample, and then the total radiated intensity from the entire sample is collected in a parabolic mirror for detection. In contrast, techniques such as near-field scanning optical microscopy or PEEM employ global excitation and local detection; the entire sample is excited with a single source and local variations are detected by a scanning probe or spatially dispersive electron optics. Because of these differences, FDTD simulations from a single exciting source (plane-wave or dipole) are insufficient to reproduce the CL response.

To clarify this point, we carried out FDTD simulations for an array of individual *z*-oriented dipoles located on a 40x40 lattice centered on the origin of the spiral. Moreover, instead of examining the near-field response like that in Figures 2a and 2b, which would not reflect the global detection of CL, we calculated the total radiated far-field intensity by integrating the magnitude of the Poynting vector across a plane 200 nm above the Ag surface after each dipole excitation. This approach, although computationally expensive, affords a more accurate physical description of STEM-CL, as it captures the effect of multiple independent near-field excitations of SPPs from a rastered electron beam and the global collection of CL. The far-field intensity plot at the OAM wavelength (660 nm) is shown in Figure 2d. Only the response near the origin was calculated due to the larger computational expense of this approach. A good qualitative match can be observed between Fig. 2c and 2d, with the vortex-like spiral arms emanating from the origin of the spiral in both figures. This distinction between the detected CL response and the direct near-field response reveals much about the utility of CL, and warrants further discussion.

*Electron Beam Interactions with Plasmonic Vortex Generators*

In electron microscopy, the likelihood of exciting a given transition is determined by the loss function, the probability per unit length of a fast electron travelling through a material to undergo energy loss for excitations at a particular energy value[33]. In many systems, the energy-loss of plasmonic materials is highly correlated to the near-field amplitude of the plasmon modes, but more generally, the loss function can be directly related to the photonic LDOS,[34]

$$\Gamma(\boldsymbol{R}, \hbar\omega) = \frac{2\pi e^2}{\hbar\omega} \tilde{\rho}_{\hat{z}}(\boldsymbol{R}, q, \hbar\omega). \qquad (2)$$

Here, $\Gamma(\mathbf{R}, \hbar\omega)$ is the loss probability per unit length, at position $\mathbf{R}$ and energy $\hbar\omega$. $\tilde{\rho}_{\hat{z}}(\mathbf{R}, q, \omega)$ is the photonic LDOS in the $\hat{z}$ direction, at position $\mathbf{R}$, wavevector $q$, and excitation energy $\hbar\omega$. Equation 2 illustrates an advantage of STEM-CL analysis of complex optical structures like plasmonic vortex generators, because it provides the ability both to detect nanoscale optical effects and to spatially map the excitation strength. The nature of the fast-electron/material interaction prevents selective excitation, but in the absence of nonlinear interactions, spectral filtering of the CL largely sidesteps this issue.

More significantly, the combination of local excitation and global detection in STEM-CL enables the detection of the plasmon phase at the nanoscale because of interference between the plasmon CL and isotropic transition radiation. As observed in Figure 1, the CL response at wavelengths far away from the OAM mode resembles linear interference fringes rather than a chiral plasmonic vortex. Similar interference patterns have been previously observed in the scanning electron microscope CL characterization of linear plasmonic gratings and attributed to interference between SPPs and transition radiation (TR) excited by fast electrons impinging on the metal surface, a phenomenon not accounted for in the near-field amplitude[35]. Transition radiation is a result of a charged particle passing between two different dielectric media. Kuttge et al. have modeled the interference between TR and SPPs as[35],

$$I_{CL} = \int_{mirror} d\Omega |A_{SPP} S(\Omega) e^{i\phi} + f_{TR}(\Omega)|^2, \quad (3)$$

where $\Omega$ is the angle of the emission, $A_{SPP}$ is the SPP amplitude, $S(\Omega)$ is the normalized in-plane far-field amplitude of the SPP after being scattered by a grating, $f_{TR}$ is the far-field amplitude of the transition radiation, and $\phi$ is the phase difference between the SPP and TR.

The impact of the TR and SPP on the detected CL of the plasmonic vortex can be seen in the simulated scan of far-field intensity in Figure 2d, because the radiation from the exciting point-dipoles in each calculation of the array give an accurate representation of TR[35]. The radiated intensity calculated in the simulation is determined by the amplitude of the Poynting vector, and hence reflect the constructive and destructive interference between the dipole and plasmonic emission define the far-field profile. The vortex is resolved in the far-field simulation but not in the near-field, demonstrating that the interference of the TR and SPP emission is the defining feature in the way CL observes plasmonic vortices. Furthermore, Equation 3 shows that, in addition to the incoherent sum of the SPP and TR intensities, the CL response contains a convolution of the plasmon amplitude and phase. Hence, the vorticity observed in the CL plasmon response directly observes the phase of a plasmonic vortex at the nanoscale, without recourse to far-field measurements of the phase of scattered or emitted light.

*Plasmon Phase Revealed Through Localized Interference*

To illustrate the influence of the plasmon phase on the detected CL signal, different geometries of vortex generators possessing different topological charges were examined using BP-filtered CL in Figure 3. The filters are all 40 nm wide and are centered at 445 nm (a-c), 513 nm (d-f), 586 nm (g-i), and 685 nm (j-l), respectively. The three structures are spiral channels with $r_0=\lambda_{SPP}$, and $d=\lambda_{SPP}$ (a, d, g, j), $2\lambda_{SPP}$ (b, e, h, k), and $3\lambda_{SPP}$ (c, f, i, l), resulting in topological charges of $\ell=1$, 2, and 3 respectively. The effect of the phase of the plasmonic vortex in the CL signal is evident for all three structures in both the 586 nm and 685 nm BP-filtered CL images. As discussed previously, and as shown in Figure 2, only the phase of the vortex shows an angular variation while its amplitude is rotationally symmetric. Thus, the presence of a vortex response in the CL implies that the plasmon phase has a strong influence on the total detected signal.

It is fruitful to take advantage of the larger field-of-view afforded by the PMT detector to analyze the plasmonic dispersion outside the spiral channel. Here, interference fringes are observed for all three structures, that result from the interference between TR and SPP emission, which are wavelength dependent but not structure dependent. The fringes are not dependent on the inter-arm spacing of the spiral because the electron beam does not excite the plasmonic vortex outside the channel. However, the fringes are dependent on the wavelength, because different SPP wavelengths accumulate different total changes in phase with respect to the TR between the points of excitation (electron probe position) and radiation (channel edge). The presence of such interference outside the spiral demonstrates that the plasmon phase modulates the detected signal. The CL profile within the spirals exhibits a strong dependence on both the dimensions of the channel as well as the wavelength, indicating both the phase structure and topological charge of the plasmonic vortex influence the interference patterns observed in Figure 3.

Access to the phase of plasmonic vortices through CL is an important observation, because generally near-field techniques employ steady-state time-averaged measurements that are insensitive to the phase of optical modes[26–28,36]. The phase structure of an optical mode is most often accessed without near-field spatial resolution via an interferogram, where the emitted signal of the mode of interest is overlapped with a separate coherent source on some adjustable offset to impart a locked phase difference to the system[1–3]. Spektor *et al*. have recently mapped phase effects in plasmonic vortex generators at the near field by using time-resolved pump-probe PEEM, where the delay between the exciting pump and the detecting probe locks the phase of the plasmonic vortex and allows it to be probed with high spatial resolution[30]. Cathodoluminescence, however, can access the phase of the plasmonic vortex without external sources or temporal resolution vortices and nanoscale chiral structures due to the combination of TR/SPP interference and spectral resolution.

The interference between TR and SPP emission provides the locked phase of an interferogram or time-resolved measurement because the TR is emitted immediately when the fast-electron impinges on the metal surface, but the SPP propagates to the channel edge before luminescing. There is a fixed distance between the point of excitation (e-beam) and the point of luminescence (channel edge) which imparts a constant phase-difference between the TR and SPP emission for each probe position. It is important to note that it is the phase convolution of TR and SPP emission that is present in the CL signal and not the direct near-field phase of plasmonic vortex, the two of which are correlated but not synonymous. The convolution is critical, because in the absence of the TR phase (*i.e.* only the SPP emission was collected) the CL profile would not possess localized sensitivity to the phase structure of the vortex.

The spectral resolution comes into play because the phase mismatch of the TR and SPP luminescence depends on both the distance from the probe to the channel and the wavevector of the SPP, and averages out across the entire dispersion. Figures 3m-o show the unfiltered CL images of the three structures in Figures 3. In these images, the total spectral response of the plasmonic vortex is collected simultaneously and integrated to show the total luminescent intensity of the vortex. There is relatively little spatial variation in the unfiltered CL profiles of the three spirals, and no vortex-like behavior is observed. Thus, for a time-averaged experiment, phase effects can only be observed with near-field spatial resolution if one has the spectral resolution to isolate narrow bands of SPP wavevectors generated by the fast-electron electrons and their interference with TR.

*Near-Field Plasmon Amplitude Signatures in Cathodoluminescence*

The simultaneous detection of the near-field plasmonic vortex amplitude in the CL signal most clearly observable in the 685 nm BP-filtered CL images in Fig. 3j-l, that is, in the wavelength bin closest to the free space OAM wavelength of 660 nm. Near the origin in all three spirals, a null region of reduced CL intensity is observed. A line profile of the CL images in Figures 3j-l is used to verify the position of the null region with respect to the origin of the spiral. Figures 4a-c show STEM-BF images of the structures with $d=\lambda_{SPP}$ (a), $2\lambda_{SPP}$ (b), and $3\lambda_{SPP}$ (c), respectively. The arm separations ($d$) in all three structures are integer multiples of the initial radius ($r_0$), and hence the near-field radial response is modulated by the Bessel function, $J(k_r r)$ where $\ell=1$ (j), 2 (k), and 3 (l), all which have a null-region at $r=0$.

To verify that the CL null region is at the origin of the spiral we examine line profiles of the 685 nm BP-filtered CL images from Figure 3j-l. The direction of the line profile is defined by the tips of the inner and outer arms of the spiral channels. Since the spiral has a winding number of $2\pi$, a line defined by the tips of the arms of the spiral should pass directly through the origin (dashed line in Figures 4a-c). Moreover, since all three structures were fabricated with an initial radius of 630 nm ($\lambda_{SPP}$), the origin (white line marked O on the line profile) should be 630 nm from the inner edge of the spiral channel along the tip-to-tip line (white line marked E on the profile).

Figure 4d shows the intensity of the 685 nm BP-filtered CL images 2 μm line profiles defined in Figures 4a-c. The line profiles begin just outside the inner arm, and are aligned such that the inner edge (E) is defined as the zero point for all three line profiles, hence the origin (O) of the spiral channel is 630 nm from the inner edge. In all three, the measured null regions lie near the predicted origin, and become more pronounced as the topological charge of the structure increases. While the null region is almost exactly at 630 nm for the 1 $\lambda_{SPP}$ structure, the 2 $\lambda_{SPP}$ and 3 $\lambda_{SPP}$ structures exhibit deviations of ~30 nm (~5 %). We attribute the error to the FIB milling of the channel, which leaves undefined edges on nanostructures due to redeposition, probe tails, and amorphization[36]. The relative agreement between the predicted and effective origins of the spiral structures demonstrate the ability of CL to detect the near-field plasmonic signatures of OAM modes and directly examine perturbations from ideal models.

*Enhanced Coupling in Same-Chirality Structures*

By experimentally characterizing the photonic LDOS, we have the capacity to analyze the near-field interactions of coupled chiral nanostructures. Figures 5a and 5b show BF-STEM images of two spiral structures (each with $d=2\ \lambda_{SPP}$). In each, a secondary spiral an order of magnitude smaller in size is milled at the origin of the larger outer spiral. For the first structure (Figure 5a), the inner spiral has the opposite chirality (OC) of the outer spiral, while for the second structure (Figure 5b), the inner structure has the same chirality (SC). To assess the coupling between the inner and outer spirals, the two structures are examined simultaneously with unfiltered PMT-CL imaging.

Figures 5c and 5d show a single unfiltered CL image of the two structures (Figure 5c) and the line profile of the CL images taken from across the origin of the two inner spirals (Figure 5d). The opposite-chirality structure is on the left, and the same-chirality structure is on the right. The CL experiment reveals that the luminescent intensity generated in the same-chirality structure is significantly greater than that produced by the opposite-chirality structure. Since the images are spectrally integrated, the phase-induced interference patterns are not observed, and the difference between the two chiral structures can therefore be directly connected to the photonic LDOS of the individual systems.

Furthermore, by collecting spectra from the inner spirals of the two systems, the spectral range of the luminescence enhancement can be observed. In Figure 5e, we compare individual spectra from the inner spiral of each pair of spiral strictures. The enhanced CL intensity in the same chirality spirals spans across a broad bandwidth in the visible regime (between 450 nm and 620 nm). The enhanced luminescence across this spectral region signals that the photonic LDOS for a broad bandwidth of plasmon modes has increased. We recall from Figure 1 that the plasmonic vortex in an isolated spiral channel was also observed across a similar bandwidth, thus reinforcing the observation of enhanced coupling between plasmonic vortices and nanoscale chiral structures.

**Conclusion**

Chiral nanostructures such as the Archimedean spiral enable direct examination of OAM at the near field, by producing optical vortex beams from features with nanoscale dimensions. By sampling the LDOS defined by the asymmetric geometry using STEM-CL and incorporating TR into the analysis, the phase and amplitude of the plasmon response of these structures can be probed directly with the strongest combination of spatial and spectral resolution available. With a spatially and spectrally resolved description of the plasmonic vortex, near-field signatures of optical orbital angular momentum modes can be detected and mapped, and effects such as enhanced coupling between chiral structures can be observed. The ability to experimentally examine near-field interactions of complex composite plasmonic effects, such as plasmonic vortices, opens the door to potential applications in molecular chirality detection and manipulation of twisted light in nanoscale structures.

**Figures**

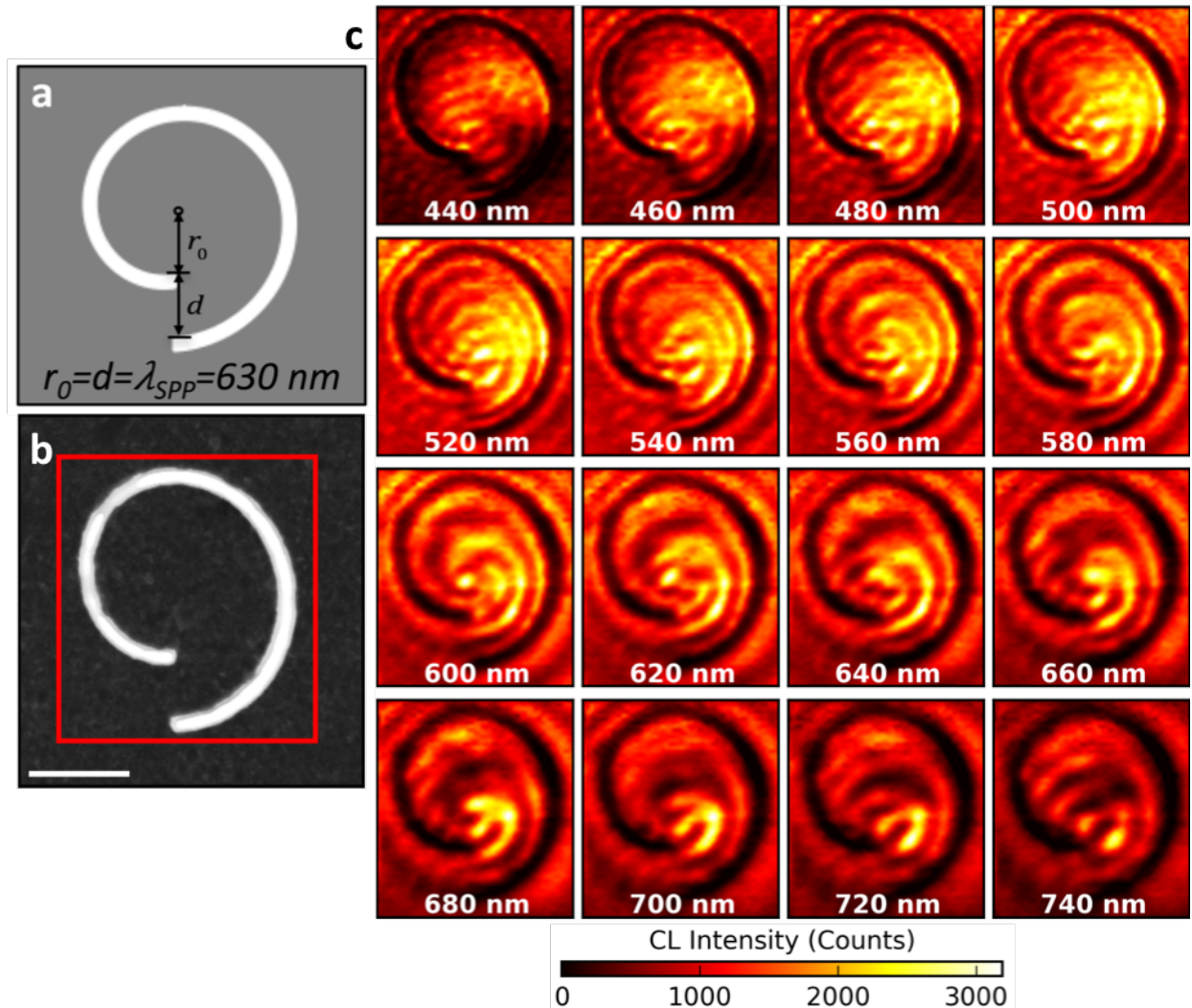

**Figure 1**. Spatially and spectrally resolved plasmonic vortices. (a) Schematic of spiral channel, with initial radius ($r_0$) equal to the arm separation ($d$) equal to the surface plasmon polariton (SPP) wavelength (630 nm). The result is a plasmonic vortex with a predicted free-space wavelength of 660 nm. (b) Bright-field (BF) scanning transmission electron microscopy (STEM) image of a spiral channel in a silver film. The red box marks the region for a spectrum image (SI). Scale bar=1 µm. (c) Cathodoluminescence (CL) SI of the spiral channel. Each frame shows a 10 nm wavelength slice of the spectrum image centered at the value on the image, showing the hyperspectral response of the vortex from 440 nm to 740 nm.

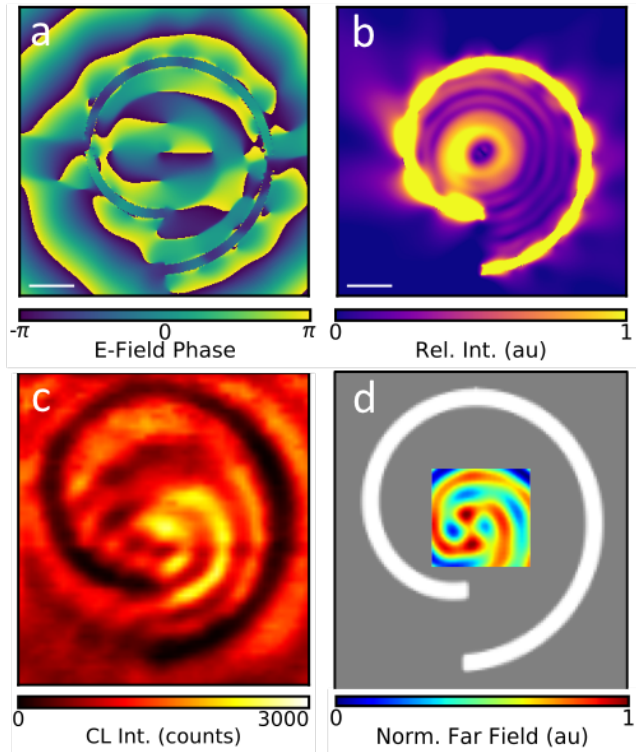

**Figure 2**. Cathodoluminescence signature of plasmonic vortex. (a,b) Finite-difference time-domain (FDTD) simulations of the phase (a) and near-field intensity (b) of a plasmonic vortex generated from a single dipole excitation at the origin of a spiral channel at the predicted free space wavelength of the vortex (660 nm). (c) CL response of plasmonic vortex at 660 nm. (d) FDTD simulation of the total intensity radiated to the far field by a 40x40 array of dipole excitations at 660 nm, showing a strong match to the detected CL signal at the vortex wavelength. Scale bar=500 nm.

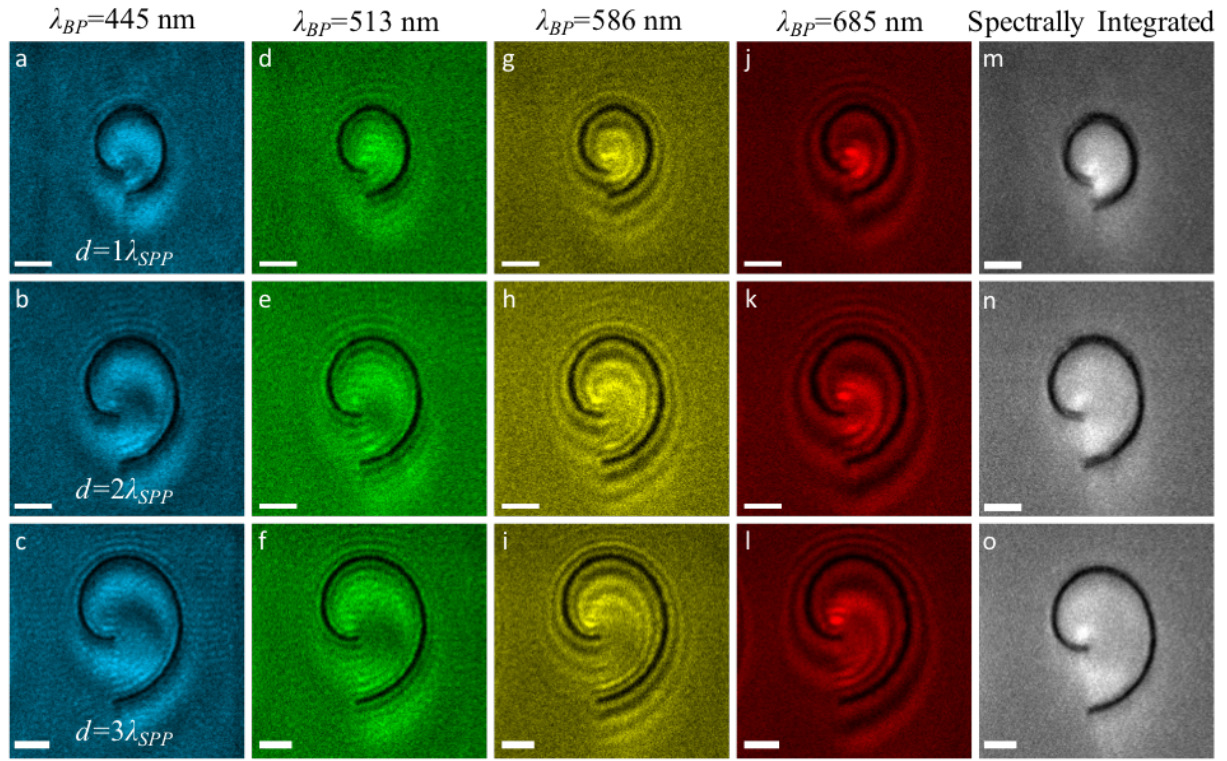

**Figure 3** Spectrally filtered plasmon cathodoluminescence maps. (a-l) Band-pass filtered (BP) CL images of spiral channels with arm separations equal to to $\lambda_{SPP}$, 2 $\lambda_{SPP}$, and 3 $\lambda_{SPP}$. The BP filters are centered at 445 nm (a-c), 513 nm (d-f), 586 (g-i), and 685 nm (j-l), and exhibit the vortex-response of the OAM mode, generated by the phase of the plasmon modes. (m-o) Unfiltered CL images of the same spiral channels, showing that the phase-induced effects are only detected with spectral selectivity, and cannot be seen in the spectrally integrated luminescence of the system. Scale bars=1 μm.

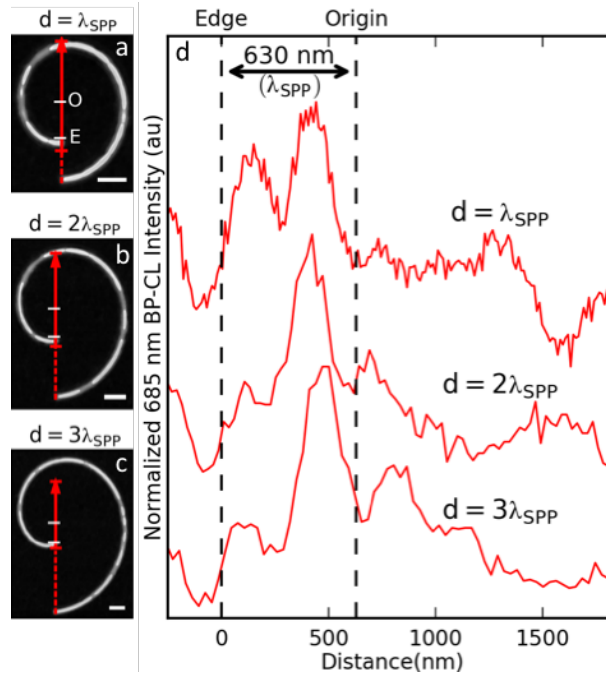

**Figure 4** Near-field amplitude signatures. (a-c) STEM-BF images of the d=$\lambda_{SPP}$ (a), 2 $\lambda_{SPP}$ (b), and 3 $\lambda_{SPP}$ (c) structures with a line profiles through the origin. Since the spiral has a winding number of $2\pi$, the origin (O) is along a line is defined by the tips of the inner and outer arm (dashed line in BF images) at a point 630 nm ($\lambda_{SPP}$) from the edge of the inner arm (E). The lines profiles begin just outside the inner arm and extend through the top of the spiral channel. (d) Line profiles of the normalized 685 nm BP-CL intensity (Figures 3j-l) along the lines defined in (a-c). Each shows a null-region centered at the origin of the spiral. Scale bars=500 nm.

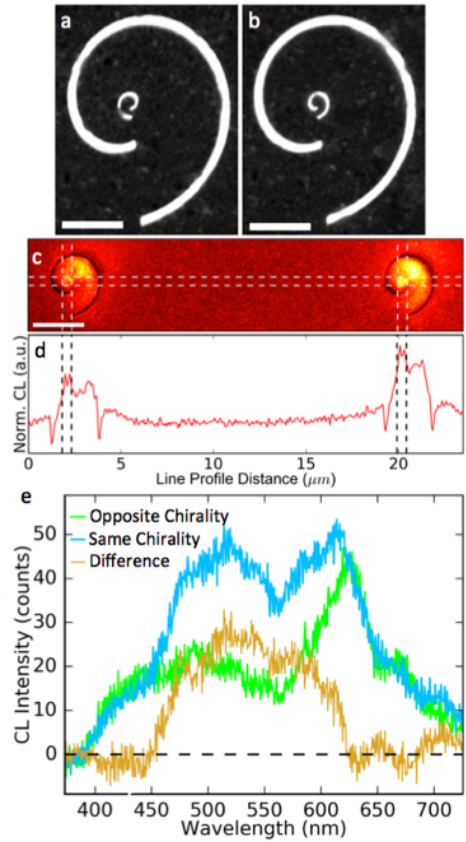

**Figure 5**. Coupling between plasmonic vortices and chiral structures. (a,b) BF-STEM image of two spirals (each with 2 $\lambda_{SPP}$ arm spacing) with a secondary spiral milled at the origin of the outer spiral, one with the opposite chirality (OC) to the outer spiral (a), the other with the same chirality (SC) as the outer spiral (b). Scale bars=1 μm. (c) Unfiltered CL images of the two structures (OC-left, SC-right), showing enhanced luminescent intensity from the SC structure. (d) Line profile across the inner spirals in (c). Scale bar=2 μm. (d) CL spectra from the center of each inner spiral, showing that the SC structure shows broad enhancement across the visible spectrum.

**Supplementary Information**
Supplementary information accompanies the manuscript on the Light: Science & Applications website (http://www.nature.com/lsa/)' at the end of the article and before the references.

**Acknowledgements**
This research was sponsored by the Laboratory Directed Research and Development Program of Oak Ridge National Laboratory, managed by UT-Battelle, LLC, for the U.S. Department of Energy (JAH, JCI, BJL, RBD) and by an appointment to the Oak Ridge National Laboratory Historically Black Colleges and Universities and Minority Education Institutions Summer Faculty Research Program (SYC). Additional support was provided by Department of Energy grant DE-FG02-09ER46554 and by the McMinn Endowment at Vanderbilt University (JAH, STP), by the Department of Energy grant DE-FG02-01ER45916 (RBD, RFH), and the Center for Nanophase Materials Sciences (CNMS), which is sponsored at ORNL by the Scientific User Facilities Division, Office of Basic Energy Sciences, U.S. Department of Energy (JAH, JCI).

Microscopy experiments were performed at the Oak Ridge National Laboratory, supported by the Department of Energy Office of Science, Basic Energy Sciences, Materials Science and Engineering Directorate (MFC). The nanospirals were patterned at CNMS, and the authors acknowledge valuable assistance from Jason Fowlkes with focused ion beam milling.

**References**

1. Marrucci, L., Manzo, C. & Paparo, D. Optical Spin-to-Orbital Angular Momentum Conversion in Inhomogeneous Anisotropic Media. *Phys. Rev. Lett.* **96,** 163905 (2006).

2. Yu, N. *et al.* Light Propagation with Phase Discontinuities: Generalized Laws of Reflection and Refraction. *Science* **334,** 333–337 (2011).

3. Cai, X. *et al.* Integrated Compact Optical Vortex Beam Emitters. *Science* **338,** 363–366 (2012).

4. Naidoo, D. *et al.* Controlled generation of higher-order Poincaré sphere beams from a laser. *Nat. Photonics* **10,** 327–332 (2016).

5. Andersen, M. F. *et al.* Quantized Rotation of Atoms from Photons with Orbital Angular Momentum. *Phys. Rev. Lett.* **97,** 170406 (2006).

6. Padgett, M. & Bowman, R. Tweezers with a twist. *Nat. Photonics* **5,** 343–348 (2011).

7. Paterson, C. Atmospheric Turbulence and Orbital Angular Momentum of Single Photons for Optical Communication. *Phys. Rev. Lett.* **94,** 153901 (2005).

8. Marino, A. M. *et al.* Delocalized Correlations in Twin Light Beams with Orbital Angular Momentum. *Phys. Rev. Lett.* **101,** 93602 (2008).

9. Wang, J. *et al.* Terabit free-space data transmission employing orbital angular momentum multiplexing. *Nat. Photonics* **6,** 488–496 (2012).

10. Tamburini, F. *et al.* Encoding many channels on the same frequency through radio vorticity: first experimental test. *New J. Phys.* **14,** 33001 (2012).

11. Bozinovic, N. *et al.* Terabit-Scale Orbital Angular Momentum Mode Division Multiplexing in Fibers. *Science* **340,** 1545–1548 (2013).

12. Alexandrescu, A., Cojoc, D. & Fabrizio, E. D. Mechanism of Angular Momentum Exchange between Molecules and Laguerre-Gaussian Beams. *Phys. Rev. Lett.* **96,** 243001 (2006).

13. Mondal, P. K., Deb, B. & Majumder, S. Angular momentum transfer in interaction of Laguerre-Gaussian beams with atoms and molecules. *Phys. Rev. A* **89,** 63418 (2014).

14. Wu, T., Wang, R. & Zhang, X. Plasmon-induced strong interaction between chiral molecules and orbital angular momentum of light. *Sci. Rep.* **5,** 18003 (2015).

15. Brullot, W., Vanbel, M. K., Swusten, T. & Verbiest, T. Resolving enantiomers using the optical angular momentum of twisted light. *Sci. Adv.* **2,** e1501349 (2016).

16. Patterson, D., Schnell, M. & Doyle, J. M. Enantiomer-specific detection of chiral molecules via microwave spectroscopy. *Nature* **497,** 475–477 (2013).

17. Su, T. *et al.* Demonstration of free space coherent optical communication using integrated silicon photonic orbital angular momentum devices. *Opt. Express* **20,** 9396–9402 (2012).

18. Strain, M. J. *et al.* Fast electrical switching of orbital angular momentum modes using ultra-compact integrated vortex emitters. *Nat. Commun.* **5,** 4856 (2014).

19. Garoli, D., Zilio, P., Gorodetski, Y., Tantussi, F. & De Angelis, F. Optical vortex beam generator at nanoscale level. *Sci. Rep.* **6,** 29547 (2016).

20. Ohno, T. & Miyanishi, S. Study of surface plasmon chirality induced by Archimedes' spiral grooves. *Opt. Express* **14,** 6285 (2006).

21. Kang, M., Chen, J., Wang, X.-L. & Wang, H.-T. Twisted vector field from an inhomogeneous and anisotropic metamaterial. *J. Opt. Soc. Am. B* **29,** 572 (2012).

22. Huang, L. *et al.* Dispersionless Phase Discontinuities for Controlling Light Propagation. *Nano Lett.* **12,** 5750–5755 (2012).

23. Gorodetski, Y., Drezet, A., Genet, C. & Ebbesen, T. W. Generating Far-Field Orbital Angular Momenta from Near-Field Optical Chirality. *Phys. Rev. Lett.* **110,** 203906 (2013).

24. Maguid, E. *et al.* Photonic spin-controlled multifunctional shared-aperture antenna array. *Science* **352,** 1202–1206 (2016).

25. Hentschel, M., Schäferling, M., Duan, X., Giessen, H. & Liu, N. Chiral plasmonics. *Sci. Adv.* **3,** e1602735 (2017).

26. Kim, H. *et al.* Synthesis and Dynamic Switching of Surface Plasmon Vortices with Plasmonic Vortex Lens. *Nano Lett.* **10,** 529–536 (2010).

27. Liu, A.-P. *et al.* Detecting orbital angular momentum through division-of-amplitude interference with a circular plasmonic lens. *Sci. Rep.* **3,** (2013).

28. Carli, M., Zilio, P., Garoli, D., Giorgis, V. & Romanato, F. Sub-wavelength confinement of the orbital angular momentum of light probed by plasmonic nanorods resonances. *Opt. Express* **22,** 26302 (2014).

29. Chen, C.-F. *et al.* Creating Optical Near-Field Orbital Angular Momentum in a Gold Metasurface. *Nano Lett.* **15,** 2746–2750 (2015).

30. Spektor, G. *et al.* Revealing the subfemtosecond dynamics of orbital angular momentum in nanoplasmonic vortices. *Science* **355,** 1187–1191 (2017).

31. Machado, F., Rivera, N., Buljan, H., Soljačić, M. & Kaminer, I. Shaping Polaritons to Reshape Selection Rules. *ArXiv161001668 Phys.* (2016).

32. Barnard, E. S., Coenen, T., Vesseur, E. J. R., Polman, A. & Brongersma, M. L. Imaging the Hidden Modes of Ultrathin Plasmonic Strip Antennas by Cathodoluminescence. *Nano Lett.* **11,** 4265–4269 (2011).

33. García de Abajo, F. J. Optical excitations in electron microscopy. *Rev. Mod. Phys.* **82,** 209–275 (2010).

34. García de Abajo, F. J. & Kociak, M. Probing the Photonic Local Density of States with Electron Energy Loss Spectroscopy. *Phys. Rev. Lett.* **100,** 106804 (2008).

35. Kuttge, M. *et al.* Local density of states, spectrum, and far-field interference of surface plasmon polaritons probed by cathodoluminescence. *Phys. Rev. B* **79,** 113405 (2009).

36. Mirhosseini, M. *et al.* Rapid generation of light beams carrying orbital angular momentum. *Opt. Express* **21,** 30196–30203 (2013).

## **Supporting Information**
Effect of Channel Width on the Plasmon Response.

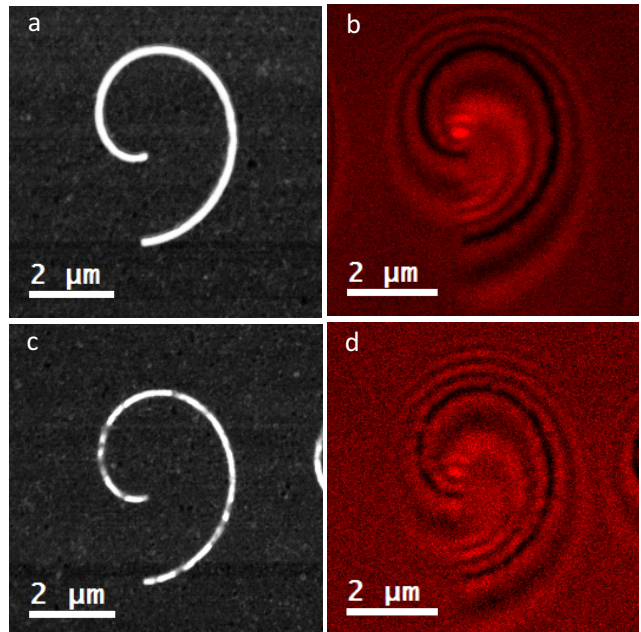

Figure S1. (a) Bright field (BF) BF scanning transmission electron microscope (STEM) image of a $d = 3\lambda_{SPP}$ with 100 nm channel from the main text. (b) 685 nm bandpass (BP) filtered photomultiplier tube

(PMT) cathodoluminescence (CL) image of the channel shown in Figure S1a (same image as Figure 3c). (c) and (d) are equaivalent images to (a) and (b) for a slit width of 75 nm.

The effect of channel width on the plasmon vortex was investigated. A bright field (BF) scanning transmission electron microscope (STEM) image (Figure S1a) and a 685 nm band-pass (BP) filtered photomultiplier tube (PMT) cathodoluminescence (CL) image (Figure S1b) of a $d=3\cdot\lambda_{SPP}$ structure with a channel width of 100 nm (same as examined in the main text). We attempted to investigate if thinner channels could achieve higher coupling efficiency of the CL to SPPs. Figures S1c and S1d show the BF and 685 nm BP PMT-CL images of $d=3\cdot\lambda_{SPP}$ structure with a 75 nm wide channel. As can be seen from the BF the channel is filled at many different locations, due to redeposition of silver during the focused ion beam milling, and the effect seen in Figure S1d is that the plasmon vortex is less pronounced. The 100 nm channel width exhibited minimal redeposition, and so was used for all subsequent experiments.

Hyperspectral response of spiral channel

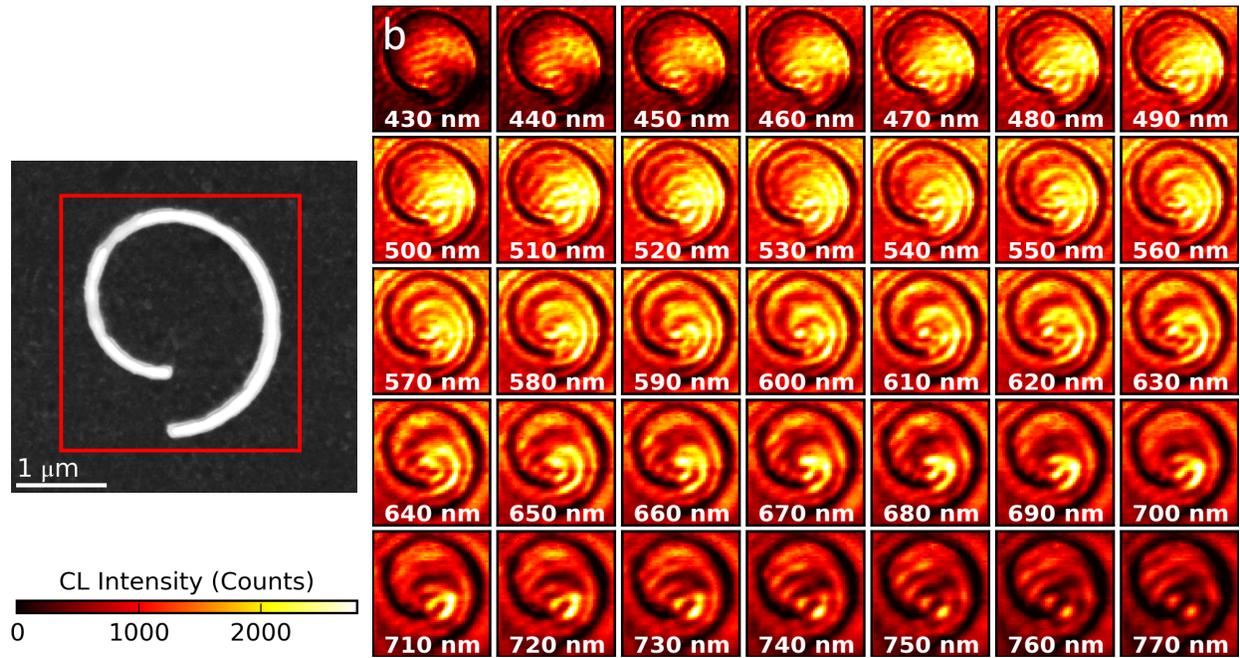

Figure S2. (a) BF-STEM image of the spiral channel shown in Figure S2a and S2b as well as Figure 2 of the main text. (b) ALL frames of the SI shown in Fig. 1 in the main text in 10 nm wavelength bins between 436 nm and 776 nm.

In the main text, (Fig. 1) a spectrum image detailing the full spectral response of a plasmonic vortex is shown in 10 nm bins acquired on a 20 nm pitch. Figure S2a shows the BF-STEM image of the spiral channel and the Figure S2b shows the SI from the main text except with all frames of the SI. It can be seen that both far below and far above the OAM wavelength of 660 nm, the CL response is dominated by linear interference fringes. However, for a bandwidth spanning 540-680 nm, the evidence of the plasmonic vortex wavefront emerges in the form of interlocking spiral arms. These spiral arms are the near-field structure of the plasmonic vortex discussed in the paper. The broad excitation of the vortex is due to (a) the fact that plasmons are generally quite broadband excitations, and (b) the relatively high thickness of the Ag film (100 nm).